# Doped Polyaniline: A Possible Anode for Organic Electronics

Akshaya K. Palai[a], N. Periasamy[b], Meghan P. Patankar, K. L. Narasimhan[c,*]

Tata Institute of Fundamental Research, Colaba, Mumbai-400 005.

ABSTRACT: Polymer based printable organic thin film transistor (OTFT) is a viable low cost alternative to amorphous silicon based thin film transistors and possesses light-weight and flexibility advantage. In this paper, we report on the hole injecting properties of doped PANI in OLED devices using it as an anode. From these results we conclude that hole doped PANI layers can be used as a low contact resistance source and drain electrode material for polymer OTFTs.

Keywords: Anode material, hole injection, OLEDs, OTFT and polymer.

## 1. Introduction

Organic electronics has emerged as a new frontier area for flexible printable electronics. Organic field effect transistors (OFET) are an important constituent with potential applications in radio frequency identification devices (RFID). OFET polymers can be used for both the active semiconductor and for source-drain contacts. Polyaniline PANI) can be synthesized using a simple and novel polymerization technique [1,2]. Heeger and co-workers have reported almost metallic transport in polyaniline with conductivities in excess of 1000 S cm$^{-1}$ [1]. PANI is hence a promising candidate for contacts for OFETs because of its easy synthesis with high conductivity and hole injecting properties.

Although there are reports on conducting polymers used as both source and drain electrodes in OTFTs [3,4], reports using doped PANI with high conductivity as a contact for source and drain for high-performance OTFT is not common. This provided the motivation for this work . In this paper, we report on hole injection properties of highly conducting PANI, doped with camphor sulfonic acid (CSA), synthesized using the procedure reported by Lee et al. [1].

## 2. Experimental details

### 2.1. Materials

Aniline (S. D. fine-chem Ltd., India) was distilled under reduced pressure prior to use. Reagent grade ammonium  persulfate, lithium fluoride (LiF) and solvents needed for the synthesis were purchased from S. D. Fine-Chem Ltd. (India). Camphor-10-sulfonic acid (CSA) (Aldrich) was used as received. Tris(8-hydroxyquinoline) aluminum (Alq$_3$) and N,N'-diphenyl-N,N'-bis(3-methylphenyl)-1,1'-biphenyl-4,4'-diamine (TPD) were obtained from Dojindo Laboratories (Japan) and Syntec GmbH (Germany) respectively. The other materials used in the device structures, 2,3,5,6-tetrafluoro-7,7',8,8'-tetracyano-p-quinodimethane (F4TCNQ), 2,9-dimethyl-4,7-diphenyl-1,10-phenanthroline (BCP) and aluminum (Al) were sourced from Aldrich and used as obtained. Indium tin oxide (ITO) on glass was purchased from Delta Technologies (U.S.A.).

### 2.2. Synthesis of Doped Polyaniline

In a typical PANI synthesis, the aqueous acidic solution of aniline (20 mmol) and chloroform (vol./vol. = 2:1) is mixed in a three necked flask. This heterogeneous biphasic mixture was cooled to 0 $^{o}$C and stirred for 30 minutes. An aqueous acidic solution of ammonium persulfate (25 mmol) was added dropwise to the above dispersion under same condition to start polymerization. The reaction mixture was filtered, washed with deionized water, acetone, deprotonated with aqueous ammonium hydroxide solution and then dried under vacuum to obtain PANI in emeraldine base (EB) form [1]. PANI-EB (2.4 mmol) was mixed with CSA (1.2 mmol) in agate mortar and pestle in a glove bag filled with nitrogen gas giving PANI-CSA [5]. High quality films of PANI-CSA as well as its blend with TPD were obtained by spin coating the material from solution in m-cresol.

## 2.3. OLED device Fabrication

Thin films of N, N'-diphenyl-N, N'-bis(3-methylphenyl) 1, 1'-biphenyl-4, 4' diamine (TPD) (40 nm), tris(8-hydroxyquinoline) aluminum ($Alq_3$) (40 nm), 2,9-dimethyl-4,7-diphenyl-1,10-phenanthroline (BCP) (3 nm) followed by bilayer of LiF (0.5 nm)-Al (30 nm) cathode were deposited in vacuum, at a base pressure of 2 x $10^{-6}$ Torr on patterned ITO coated glass substrates or spin coated with doped PANI. The doped PANI was spin coated at a spin speed of 1500 rpm for 1 minute and then dried at 100 $^o$C for 10 minutes inside a glove box. The device structures were fabricated with configuration of

    a.  ITO/TPD/$Alq_3$/BCP/LiF/Al

    b.  ITO/PANI-CSA/$Alq_3$/BCP/LiF/Al

    c.  ITO/PANI-CSA/TPD/$Alq_3$/BCP/LiF/Al

    d.  ITO/PANI-CSA:TPD (w/w = 1:1)/TPD/$Alq_3$/BCP/LiF/Al and

    e.  ITO/F4TCNQ/TPD/$Alq_3$/BCP/LiF/Al

The active area of the device was 2 mm$^2$. Devices (a) and (e) are standard OLED devices based on TPD/$Alq_3$. Device "e" has a F4TCNQ layer which facilitates hole injection into TPD and lowers the operating voltage. These two devices serve as a benchmark for comparison with devices "b-d". Devices "b-d" were made using doped PANI as the anode. Current, voltage and electro-luminescence (EL) characteristics were measured using a Keithley electrometer and DVM. The details are reported elsewhere [6].

## 3. Result and discussion

### 3.1 DC Conductivity

The dc conductivity of CSA doped PANI films was measured by spin coating the doped PANI on pre-patterned ITO substrates for planar measurements. The thickness of the films was 100 nm. The current-voltage characteristics was ohmic. The typical conductivity of the CSA doped PANI films was 100 S cm$^{-1}$.

### 3.2 OLED device properties

Figure 1 is a plot of the current density (J) versus voltage (V) for devices a-e. Figure 2 shows the electroluminescence (EL) intensity as a function of applied voltage for the PANI devices. Figure 3 shows the EL intensity of device as a function of applied voltage for device "a" and "e" respectively. Figure 4 shows the EL intensity as a function of the current density for devices a-e. We now discuss these results.

The standard Alq$_3$ device "a" has a large turn-on voltage due to a large barrier to hole injection between TPD and ITO. The introduction of a 3 nm F4TCNQ layer (device "e") dramatically reduces the barrier to hole injection as the TPD layer gets hole doped by diffusion of F4TCNQ [7]. The doped PANI and doped PANI:TPD (devices c and d respectively) have a significantly lower turn-on voltage than device "a". This is also true for the CSA doped PANI layer in direct contact with $Alq_3$. These results confirm that doped PANI is a good hole injector.

In figure 2 we see that the EL of the doped PANI devices also turns on at a lower voltage than the standard $TPD/Alq_3$ device (device "a"). The EL of device "e" turns on at a lower voltage than device "a" due to more efficient hole injection. This is seen clearly in figure 3. Figure 4 shows the EL intensity (L)-Current density (J) plots for the standard device and the doped PANI devices. The L-J slope is much smaller for the doped PANI devices than that for the standard $TPD/Alq_3$ device (device "a"). The L-J slope is similar for devices "a" and "e". A high L-J slope indicates balanced electron–hole injection. In device "a", this is facilitated by the large barrier to hole injection from the ITO and the presence of the BCP layer which is a hole blocking layer. The L-J slope for device "e" is slightly smaller than that for device "a" primarily due to the fact that the marginally hole doped TPD improves the hole injection efficiency and is responsible for the lower operating voltage. The small L-J slope seen for devices with doped PANI as an anode is consistent with the fact that the current in the doped PANI devices is dominated by one carrier. Since the PANI layer is hole doped this suggests that the current in devices "b-d" are hole dominated. These results are consistent with the fact that doped PANI layers are good hole injectors.

These results are of great interest for printable polymer electronics. For printed organic TFT, PEDOT:PSS is often used as the hole contact for source and drain electrodes. Long term degradation of PEDOT:PSS devices have been reported [8]. The results reported here demonstrate that doped PANI devices are a viable replacement for PEDOT:PSS for source (drain) electrodes in TFT. The source (drain) electrodes can be printed using doped PANI blended with TPD. This will help in reducing series resistance and also enable transistors to operate at higher frequency and enable the fabrication of short channel transistors in a manufacturing environment. The doped PANI layer can be used for both small molecule and polymer FET's as contact layers.

## 4. Conclusion

In conclusion, we have demonstrated that doped PANI, blended with TPD is a good hole injector and a substitute for PEDOT:PSS. In particular it can be used as source and drain contacts in printed hole conducting OFETs.

**References**

[1] K. Lee, S. Cho, S.H. Park, A.J. Heeger, C.-W. Lee, S.-H. Lee, Nature 441 (2006) 65.

[2] S.-H. Lee, D.-H. Lee, K. Lee, C.-W. Lee, Adv. Funct. Mater. 15 (2005) 1495.

[3] T.G. Backlund, H.G.O. Sandberg, R. Osterbacka, H. Stubb, T. Makela, S. Jussila, Synth. Met. 148 (2005) 87.

[4] H. Rost, J. Ficker, J.S. Alonso, L. Leenders, I. McCulloch, Synth. Met. 145 (2004) 83.

[5] K. Lee, A.J. Heeger, Y. Cao, Phys. Rev. B 48 (1993) 14884.

[6] P.K. Nayak, N. Agarwal, N. Periasamy, M.P. Patankar, K.L. Narasimhan, Synth. Met. 160 (2010) 722.

[7] F. Ali, N. Periasamy, M.P. Patankar, K.L. Narasimhan, J. Appl. Phys. 110 (2011) 044507.

[8] K. Kawano, R. Pacios, D. Poplavskyy, J. Nelson, D.D.C. Bradley, J.R. Durrant, Sol. Energy Mater. Sol. Cells 90 (2006) 3520.

# Figure Captions

1. **Fig. 1** Current density vs Voltage curve of the devices (■) ITO/TPD/Alq$_3$/BCP/LiF/Al, (●) ITO/PANI-CSA/Alq$_3$/BCP/LiF/Al, (▲) ITO/PANI-CSA/TPD/Alq$_3$/BCP/LiF/Al, (▼)ITO/PANI-CSA:TPD/TPD/Alq$_3$/BCP/LiF/Al and (♦) ITO/F4TCNQ/TPD/Alq$_3$/BCP/LiF/Al respectively.

2. **Fig. 2** EL intensity vs Voltage curve of the devices (■) ITO/TPD/Alq$_3$/BCP/LiF/Al, (●) ITO/PANI-CSA/Alq$_3$/BCP/LiF/Al, (▲) ITO/PANI-CSA/TPD/Alq$_3$/BCP/LiF/Al, (▼) ITO/PANI-CSA:TPD/TPD/Alq$_3$/BCP/LiF/Al and (♦) ITO/F4TCNQ/TPD/Alq$_3$/BCP/LiF/Al respectively.

3. **Fig. 3** EL intensity vs Voltage curve of the devices (■) ITO/TPD/Alq$_3$/BCP/LiF/Al and (♦) ITO/F4TCNQ/TPD/Alq$_3$/BCP/LiF/Al respectively.

4. **Fig.4** EL intensity vs Current density curve of the devices (■) ITO/TPD/Alq$_3$/BCP/LiF/Al, (●) ITO/PANI-CSA/Alq$_3$/BCP/LiF/Al, (▲) ITO/PANI-CSA/TPD/Alq$_3$/BCP/LiF/Al, (▼) ITO/PANI-CSA:TPD/TPD/Alq$_3$/BCP/LiF/Al and (♦) ITO/F4TCNQ/TPD/Alq$_3$/BCP/LiF/Al respectively.

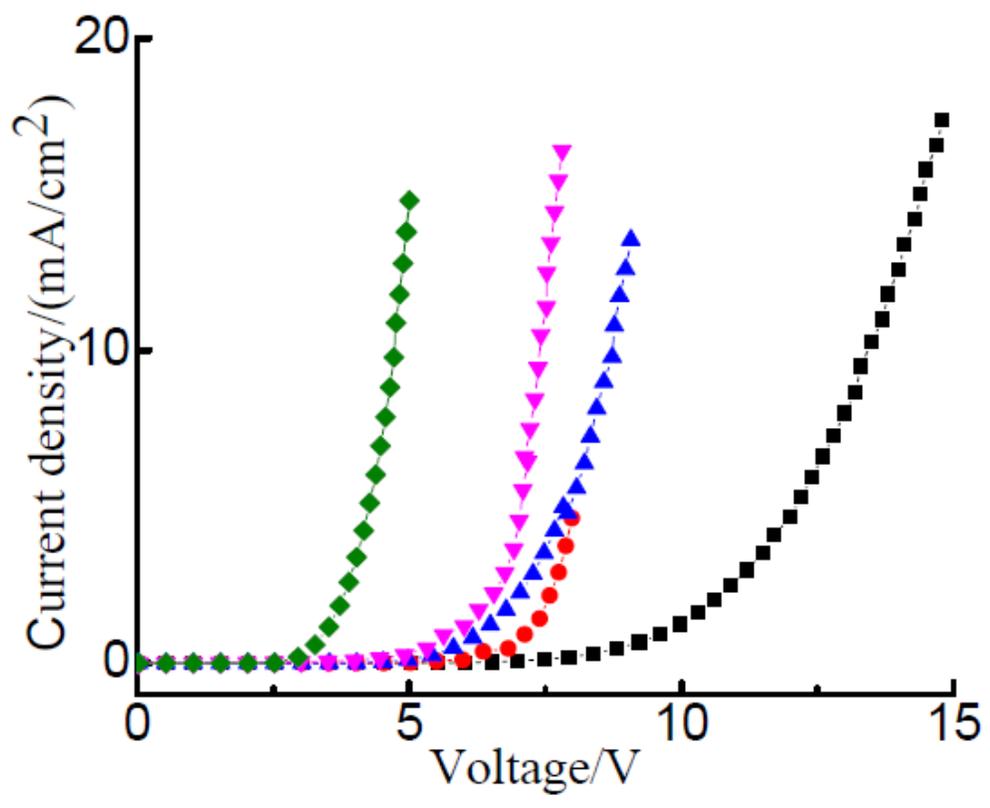

Figure 1

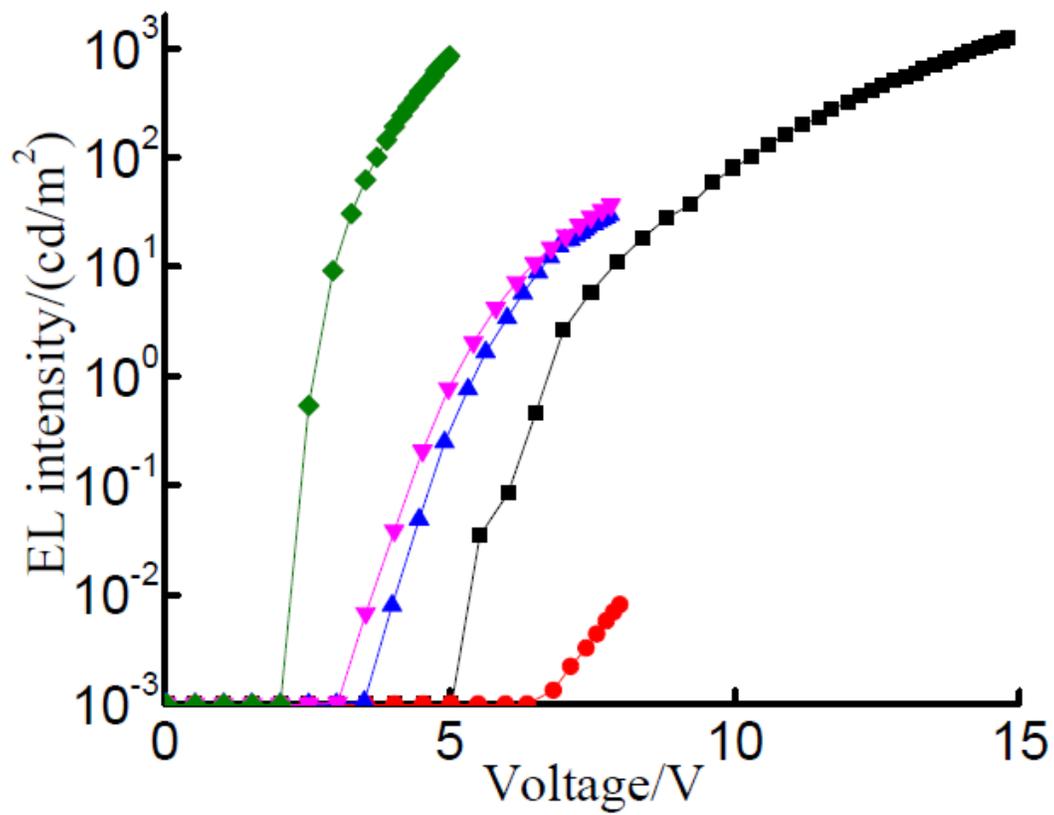

Figure 2

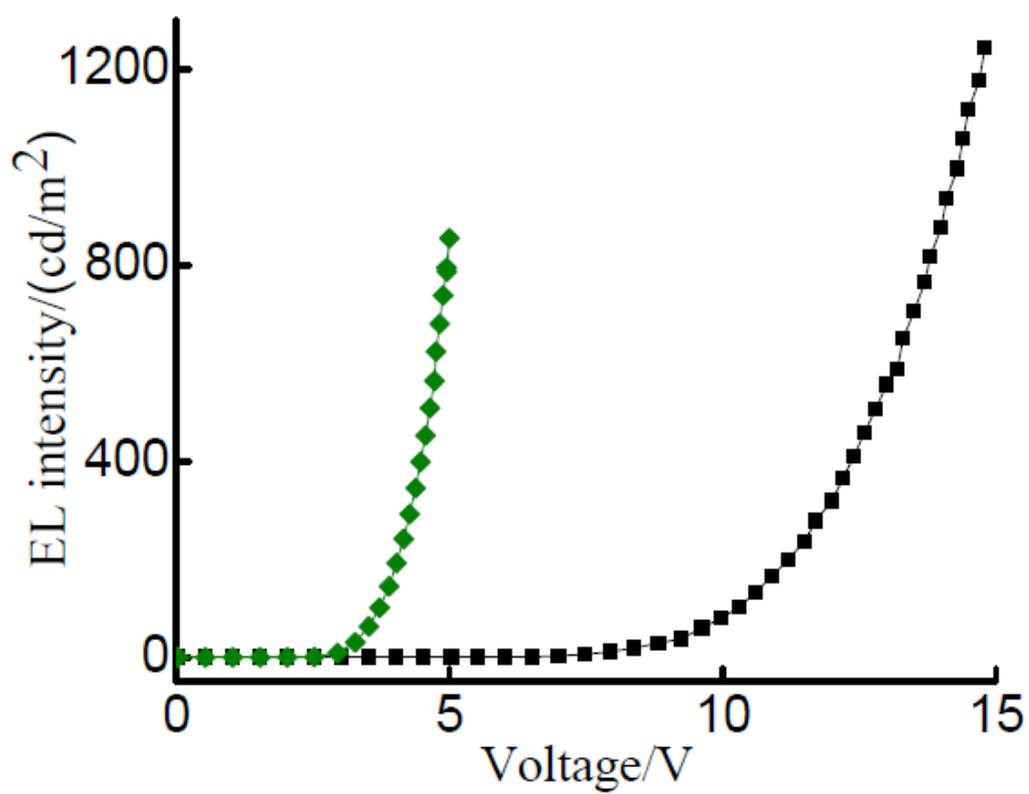

Figure 3

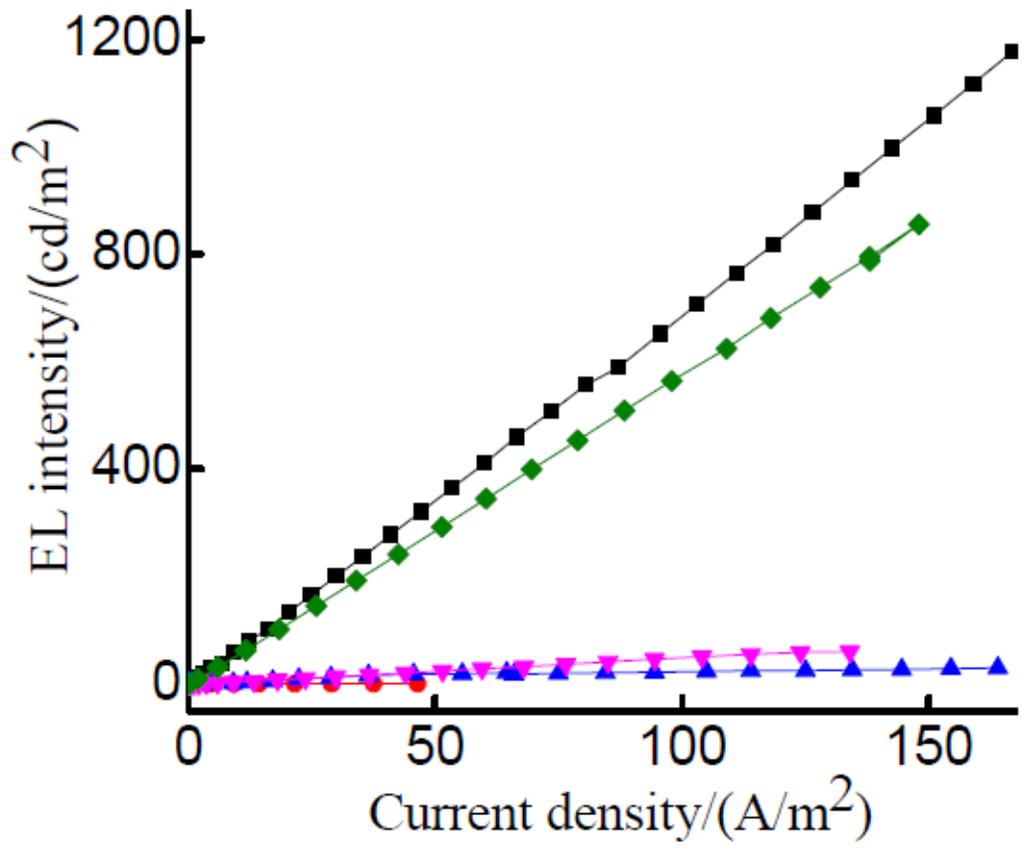

Figure 4